# Strong-field molecular alignment mediated by nonadiabatic charge localization

D.A. Romanov[1,3] and R.J. Levis[2,3]

[1]Department of Physics, Temple University, Philadelphia, PA 19122
[2]Department of Chemistry, Temple University, Philadelphia, PA 19122
[3]Center for Advanced Photonics Research, Temple University, Philadelphia, PA, 19122

**Abstract**

A new mode of effective interaction of molecular rotational degrees of freedom with an intense, nonresonant, ultrashort laser pulse is explored. Transient nonadiabatic charge redistribution (TNCR) in larger molecules or molecular ions causes impulsive-torque interaction that replaces the traditional mechanism of molecular alignment based on perturbative interaction of the laser field with electronic subsystem as manifested in linear anisotropic polarizability or hyperpolarizability. We explore this new alignment mechanism on a popular generic model of a tight-binding diatomic molecule. We consider the case of rotational wavepacket formation when a molecule is initially in the ground rotational state. The rotational wavepacket emerging from the TNCR interaction consists of states with higher rotational quantum numbers, in comparison with the anisotropic-polarizability case, and the after-pulse alignment oscillations are out-of-phase with those resulting from the traditional interaction. The TNCR interaction mode is expected to play a major role when a strong laser field actually causes extensive nonresonant excitation and/or ionization of a molecule.

**Introduction**

Laser-induced molecular alignment occurs when nonspherical molecules are exposed to a short, intense, linearly polarized laser pulse.[1] Of particular interest is non-adiabatic alignment, caused by a nonresonant interaction with a pulse of duration much shorter than the rotational period of the molecule. In this case, the rotational kick received by the molecules leads to alignment of the molecular ensemble and a series of periodic rotational revivals in the wake of the laser pulse, which are attenuated by coherence-loss mechanisms.[2] This field-free alignment phenomenon has found extensive use in various experiments addressing anisotropic characteristics of individual molecules, such as photoelectron angular distribution,[3-5] Coulomb explosion imaging,[6] control of molecular scattering,[7] and high-harmonic generation with applications to attosecond physics[8-10] and molecular tomography.[11,12]

Laser-induced molecular alignment is a powerful method for controlling the physical and optical properties of a gas-phase molecular medium, which has been used for modifying the propagation dynamics of subsequent laser pulses.[13-15] Molecular alignment has been applied to phase modulation of an optical pulse,[16,17] spectral interferometry,[18] spatial focusing/defocusing,[19,20] transient birefringence,[21-23] optical wave-guiding,[24] and even control of unimolecular chemical reactions.[25,26]

In a classical picture, the linearly polarized laser pulse gives the molecule an impulsive torque that drives the molecular axis of maximum polarizability towards the laser polarization axis. The magnitude of this torque depends on the angle between these two axes, and thus the after-pulse evolution results in alignment of the molecular ensemble. In a quantum-mechanical picture, the interaction prepares a coherent superposition of rotational states (the rotational wavepacket) that undergoes field-free evolution after the pulse. The components of the wavepacket are at first in phase and then experience dephasing due to the spread in rotational energies, followed by repeated rephasing at integer and possibly fractional multiples of the mean rotational period. The condition for effective excitation of the rotational wavepacket is that the exciting laser pulse should be shorter than the rotational period, $\pi\hbar/B_e$ where $\hbar$ is Planck's constant and $B_e$ is the rotational constant in energy units (typically, $B_e \sim 10^{-4}$ eV). The alignment will recur periodically as long as the excited states in the rotational wavepacket remain phase-locked, that is, until loss of rotational coherence due to collisions.

In all the mentioned applications of impulsive alignment, the action of the laser pulse on the rotational degrees of freedom of a molecule is mediated by nonresonant interaction with the molecular electronic system, and the latter is treated perturbatively based on the assumption that $\left(\mu\mathcal{E}_0/|\hbar\omega_c - \Delta E|\right) << 1$ where $\mu$ is the characteristic dipole matrix element between the ground and excited electronic states, $\Delta E$ is the energy distance between those states, $\omega_c$ is the laser carrier frequency, and $\mathcal{E}_0$ is the laser field magnitude (we will confine ourselves with cases of symmetric-top molecules carrying no permanent dipole). In the traditional approach, only the lowest, second-order perturbation is taken into account, and then the alignment-causing effective interaction Hamiltonian takes the form, $\hat{H}_{int}(\theta,t) = -(1/4)(\alpha_\| - \alpha_\perp)\mathcal{E}_0^2 f^2(t)\cos^2\theta$, where $\alpha_\|$ and $\alpha_\perp$ are the longitudinal and transverse components of the polarizability tensor, $\theta$ is the angle between the molecular axis and the direction of the laser polarization, while $\mathcal{E}_0$ and $f(t)$ are the electric field magnitude and the envelope function of the linearly-polarized laser pulse with carrier frequency $\omega_c$ satisfying the criterion, $\omega_c >> |\alpha_\| - \alpha_\perp|\mathcal{E}_0^2/\hbar$. At relatively weak electric fields, when $|\alpha_\| - \alpha_\perp|\mathcal{E}_0^2 < \hbar^2/2I$, where $I$ is the molecular moment of inertia, the action of $\hat{H}_{int}(\theta,t)$ on the rotational degrees of freedom can also be considered as a perturbation, and then it causes transitions between the angular momentum eigenstates with $\Delta l = \pm 2$ and $\Delta m = 0$. For stronger electric fields, the rotational Schrödinger equation with the interaction Hamiltonian $\hat{H}_{int}(\theta,t)$ needs to be solved numerically.[27] In this situation, the mediated action of the laser pulse on the molecular rotational degrees of freedom is treated non-perturbatively, while the response of the electronic degrees of freedom still allows for a perturbative treatment. For yet stronger values of $\mathcal{E}_0$, higher-order perturbations to the electronic system may be taken into account, resulting in quartic (in $\mathcal{E}_0$) corrections to $\hat{H}_{int}(\theta,t)$, related to the components of second-hyperpolarizability tensor $\gamma_{ijkl}$ (these higher-order perturbations are apparently more important in the case of asymmetric molecules where first hyperpolarizability $\beta_{ikl}$ is engaged).[28-33]

However, when the laser electric field becomes so strong that $\mu\mathcal{E}_0 \sim \Delta E$, the perturbation series for the electronic response does not converge any more, and thus incorporating higher-order perturbations in $\hat{H}_{int}(\theta,t)$ will not provide a reliable description of the effective interaction of the laser pulse with the rotational degrees of freedom. Physically, this means that the strong laser field causes essential restructuring of the electron system, which cannot be addressed via successive perturbative approximations. This situation may occur in larger molecules or molecular ions, where distances between electronic energy levels are relatively small. Another likely possibility is the situation when a molecule is driven by the laser field into the excited state manifold on the way to ionization.[34,35] For instance, when $\Delta E \sim 2$ eV and $\mu \sim 1\, e \cdot \text{Å}$, the laser intensity of $\sim 5 \cdot 10^{13}$ W/cm$^2$ makes it already for $\mu\mathcal{E}_0 \sim \Delta E$. When the strong field thus becomes capable of essentially modifying the electronic system during the laser pulse, it remains an open question what happens with the effective interaction of this laser pulse with the rotational degrees of freedom of the molecule. In this publication, we address one important aspect of such engagement.

Strong oscillating fields are known to cause nonadiabatic redistribution of the electron charge in the molecule and even induce transient charge localization.[36-38] (Note that in this context the term "nonadiabatic" relates to the electronic response and should not be confused with the term "nonadiabatic alignment", which means molecular alignment following a laser pulse of a duration much shorter than the molecular rotational period and which is called here "impulsive alignment" to avoid possible confusion.) This effect has a long and venerable history. It was first predicted theoretically for an archetypical model of a particle in a symmetric double well, driven by a monochromatic classical force,[39] and for a more general driven two-level system,[40] and termed coherent destruction of tunneling. Later, effects of this kind were found and utilized in a vast multitude of molecular and solid-state systems,[41-43] ranging from individual atoms in external potentials,[42,44] to strongly driven qubits in Josephson circuits[45] and optical traps,[46] to strongly-driven spin control for spintronics applications,[47] and even to coupled optical waveguides.[48] The nonadiabatic charge-redistribution effects under strong laser driving were found to be especially pronounced in larger molecules and in molecular ions.[49,50] As these effects depend critically on the amplitude of the electric potential variation across the molecule placed in the laser field, they should be sensitive to the molecular orientation with respect to the laser polarization. The goal of this communication is to reveal manifestations of nonadiabatic effects on the rotational degrees of freedom, modifications of the torque exerted on the molecule, and substantial changes in the composition of the resulting rotational wavepacket and the dynamics of field-free alignment.

Keeping the in line with the general approach to nonadibatic localization, we use the conceptual two-site model,[41] which in our context represents a homonuclear diatomic molecule with a single active electron in the tight-binding approximation.[51] The electron energy in an isolated atom is taken as a reference point, and the electron tunneling between the two sites leads to the level splitting, so that the energy of the ground molecular state is $E_g = -|V|$ and the energy of the excited state is $E_e = |V|$, where $V$ is the tunneling matrix element. When an electric field $\mathcal{E}_0$ is applied along the molecular axis, these two energy levels are shifted as $E_{g,e}(\mathcal{E}_0) = \mp\sqrt{V^2 + (e\mathcal{E}_0 R/2)^2}$, where $R$ is the internuclear distance. If the electric field is weak, $|\mathcal{E}_0| \ll |2V/eR|$, the expansion of $E_{g,e}(\mathcal{E}_0)$ over the small parameter $|e\mathcal{E}_0 R/2V| \ll 1$ produces the longitudinal polarizabilities of these two states as $(\alpha_\parallel)_{g,e} = \mp(eR/2)^2/|V|$ (assuming $|\alpha_\perp|_{g,e} \ll |\alpha_\parallel|_{g,e}$, the transverse polarizabilities are henceforth neglected). If the electric field is applied at an angle $\theta$ with respect to the molecular axis, the energy levels are shifted by the longitudinal component of this field: $E_{g,e} = \mp\sqrt{V^2 + (eR/2)^2 \mathcal{E}_0^2 \cos^2\theta}$. The two electronic energy levels as functions of the electric field strength and the angle $\theta$ are presented in Figure 1, where the energy is in the units of $|V|$ and the electric field in the units of $2|V|/eR$. The angular dependence of the energy levels forms the effective potential energy curves for rotational motion of the molecule. As seen, at any given value of $\mathcal{E}_0$, the effective potential energy for the ground electronic state has its minimum at $\theta = 0(\pi)$, while the effective potential energy for the excited electronic state has its minimum at $\theta = \pm\pi/2$. For the weak-field laser pulse, $\mathcal{E}(t) = \mathcal{E}_0 f(t)\cos(\omega_c t)$, using the above-mentioned polarizability values $(\alpha_\parallel)_{g,e}$ and averaging over the laser cycle leads to the effective time-dependent rotational interaction Hamiltonians, $\left(\hat{H}_{int}\right)_{g,e} = \mp(1/4)\left((eR/2)^2/|V|\right)\mathcal{E}_0^2 f^2(t)\cos^2\theta$, which provide the alignment rotational kick for a molecule in the ground state and the anti-alignment kick for a molecule in the excited state. One might expect that when the laser field is no longer weak ($|e\mathcal{E}_0 R/V| \sim 1$) and thus the polarizability-based approach is no longer valid, the strong-field version of the effective Hamiltonian will become $\left(\hat{H}_{int}\right)_{g,e} = \mp\sqrt{V^2 + (eR/2)^2 \mathcal{E}_0^2 f^2(t)\cos^2\theta}$.

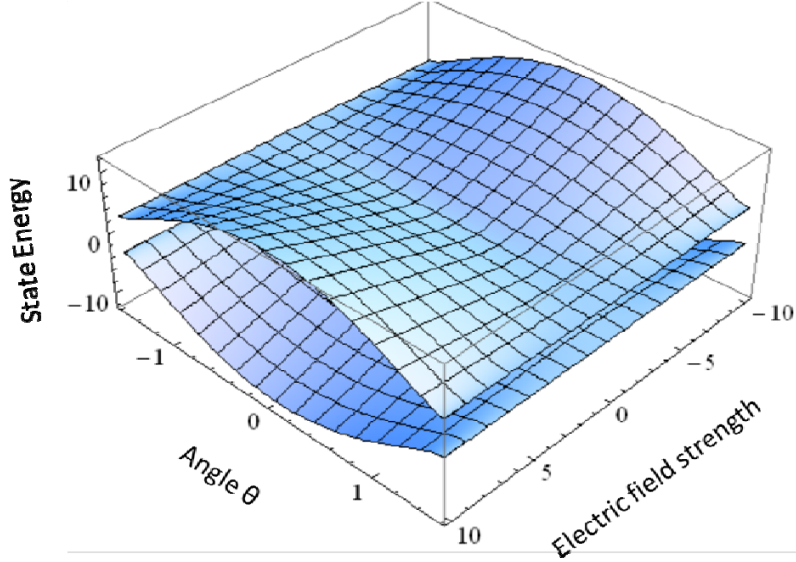

**Figure 1**. Potential energy curves for molecular rotation as depending on the strength of the external electric field. The energy is in the units of $|V|$, the electric field in the units of $2|V|/eR$. The lower surface corresponds to the ground electronic state; the upper surface, to the excited electronic state. The surfaces are shifted away from each other for clarity.

However, this weak-field interaction Hamiltonian and its strong-field generalization are both based on the implicit assumption that the molecular electronic system responds instantaneously to the oscillations of the laser field. If this is not the case, that is, if either of the conditions, $\omega_c \ll |V|/\hbar$ and $eR\mathcal{E}_0\omega_c \ll V^2/\hbar$, is not satisfied, the averaging over the carrier-frequency oscillations of the laser field produces a different effective rotational Hamiltonian, which determines an alternative dependence of the alignment kick on the molecular parameters and the laser pulse characteristics. We will explore the transition from the anisotropic-molecular-polarizability alignment mechanism to a new mode of the effective-torque interaction, which is related to nonadiabatic electron localization.

**The model**

To concentrate on general aspects of the effects the nonadiabatic electron dynamics can have on the molecular rotation, we consider a generic model of a homonuclear diatomic molecule in the single active electron approximation (each of the two nuclear cores carries positive charge of $e/2$, thus maintaining electric neutrality of the system). The molecule is subjected to a linearly polarized laser pulse with the electric field $\hat{\mathbf{e}}\mathcal{E}(t) = \hat{\mathbf{e}}\mathcal{E}_0 f(t)\cos(\omega_c t)$, where $\hat{\mathbf{e}}$ is the unit vector in the polarization direction, $\mathcal{E}_0$ is the pulse amplitude, $\omega_c$ is the carrier frequency, and $f(t)$ is the slow-varying envelope function, normalized to unity. The pulse-duration time, $\tau$, as determined by $f(t)$, is much larger than the laser cycle period, $2\pi/\omega_c$. The total molecular Hamiltonian is a function of the nuclear coordinates, $\mathbf{R}_1$ and $\mathbf{R}_2$, and the electronic coordinate, $\mathbf{r}$. The Hamiltonian is comprised of the electronic part, $\hat{H}_e(\mathbf{R}_1, \mathbf{R}_2, \mathbf{r}) = -(\hbar^2/(2m_e))\nabla_r^2 + V(|\mathbf{r}-\mathbf{R}_1|) + V(|\mathbf{r}-\mathbf{R}_2|)$, the nuclear part, $\hat{H}_n(\mathbf{R}_1, \mathbf{R}_2, \mathbf{r}) = -(\hbar^2/(2M))(\nabla_{R_1}^2 + \nabla_{R_2}^2) + V_n(|\mathbf{R}_1 - \mathbf{R}_2|)$, and the interaction part, $\hat{H}_{int}(\mathbf{r}, t) = -e(\mathbf{r}\cdot\hat{\mathbf{e}})\mathcal{E}(t) + e\mathcal{E}(t)(1/2)((\mathbf{R}_1\cdot\hat{\mathbf{e}}) + (\mathbf{R}_2\cdot\hat{\mathbf{e}}))$. Within the Born-Oppenheimer approximation, the tight-binding molecular wavefunction is represented as,

$$\Psi(\mathbf{R}_1, \mathbf{R}_2, \mathbf{r}, t) = a_1(\mathbf{R}_1, \mathbf{R}_2, t)\psi_e(\mathbf{r} - \mathbf{R}_1)\exp(-it E_0/\hbar) + a_2(\mathbf{R}_1, \mathbf{R}_2, t)\psi_e(\mathbf{r} - \mathbf{R}_2)\exp(-it E_0/\hbar),$$

where the site-wise electron functions $\psi_e(\mathbf{r})$ are the ground-state solutions of the local stationary Schrödinger equations with the potential $V(r)$, while the coefficients $a_1(\mathbf{R}_1, \mathbf{R}_2, t)$ and $a_2(\mathbf{R}_1, \mathbf{R}_2, t)$ constitute the effective two-component nuclear wavefunction, $\mathbf{a}(\mathbf{R}_1, \mathbf{R}_2, t)$, which satisfies the equations,

$$i\hbar\frac{\partial}{\partial t}\mathbf{a}(\mathbf{R}_1, \mathbf{R}_2, t) = \left(\hat{H}_n(\mathbf{R}_1, \mathbf{R}_2) + V_2(|\mathbf{R}_2 - \mathbf{R}_1|)\right)\hat{\sigma}_0 \mathbf{a}(\mathbf{R}_1, \mathbf{R}_2, t) + \\ V_1(|\mathbf{R}_2 - \mathbf{R}_1|)\hat{\sigma}_x \mathbf{a}(\mathbf{R}_1, \mathbf{R}_2, t) - \frac{e\mathcal{E}(t)}{2}\left(\hat{\mathbf{e}} \cdot (\mathbf{R}_1 - \mathbf{R}_2)\right)\hat{\sigma}_z \mathbf{a}(\mathbf{R}_1, \mathbf{R}_2, t) \quad (1)$$

Here, $\hat{\sigma}_x$, $\hat{\sigma}_z$, and $\hat{\sigma}_0$ are the Pauli matrices, and the effective potentials are determined by the overlap integrals, $V_1(|\mathbf{R}_2 - \mathbf{R}_1|) = \int d^3\mathbf{r}\, \psi_e^*(\mathbf{r} - \mathbf{R}_1)\psi_e(\mathbf{r} - \mathbf{R}_2)V(|\mathbf{r} - \mathbf{R}_1|)$ and $V_2(|\mathbf{R}_2 - \mathbf{R}_1|) = \int d^3\mathbf{r}\, \psi_e^*(\mathbf{r} - \mathbf{R}_1)\psi_e(\mathbf{r} - \mathbf{R}_1)V(|\mathbf{r} - \mathbf{R}_2|)$. (We assume that the $\psi_e(\mathbf{r})$ state of an isolated atom has no permanent dipole moment.) We separate the nuclear center-of-mass motion in Eq. (1) by introducing new variables, $\mathbf{R} = \mathbf{R}_1 - \mathbf{R}_2$ and $\mathbf{R}_+ = (\mathbf{R}_1 + \mathbf{R}_2)/2$, and representing the expected solution to Eq. (1) in the form: $\mathbf{a}(\mathbf{R}_+, \mathbf{R}, t) = \exp\left((i/\hbar)\left(\mathbf{P}\cdot\mathbf{R}_+ - (P^2/(4M))t\right)\right)\tilde{\mathbf{a}}(\mathbf{R}, t)$, where $\mathbf{P}$ is the center-of-mass momentum. Then, the equations for $\tilde{\mathbf{a}}(\mathbf{R}, t)$ read:

$$i\hbar\frac{\partial}{\partial t}\tilde{\mathbf{a}}(\mathbf{R}, t) = \left(-\frac{\hbar^2}{M}\nabla_R^2 + V_c(R)\right)\tilde{\mathbf{a}}(\mathbf{R}, t) - \frac{e\mathcal{E}(t)}{2}(\hat{\mathbf{e}}\cdot\mathbf{R})\hat{\sigma}_z\tilde{\mathbf{a}}(\mathbf{R}, t) + V_1(R)\hat{\sigma}_x\tilde{\mathbf{a}}(\mathbf{R}, t) \quad (2)$$

where $V_c(R) = V_n(R) + V_2(R)$.

For the sake of simplicity, we assume that the first term in the right-hand side of Eq. (2) provides a rigid-rotor-type configuration with some equilibrium internuclear distance $R_g$. Then, the effect of the laser-molecule interaction is determined by three competing energy scales: the rotational quantum, $\hbar^2/MR_g^2$, the tunneling level splitting, $V_1(R_g)$, and the level mismatch amplitude, $e\mathcal{E}_0 R_g$. In the absence of the laser field ($\mathcal{E}_0=0$) the system of equations is readily diagonalized by the transformation $\tilde{\mathbf{b}}(x,t') = (\hat{\sigma}_z + \hat{\sigma}_x)\tilde{\mathbf{a}}(x,t')$, Then, if $e\mathcal{E}_0 R_g \ll V_1(R_g)$, the second term in the right-hand side of Eq. (2) is treated as a perturbation, which leads to the effective rotational interaction Hamiltonians in the above-mentioned form $(\hat{H}_{int})_{g,e} = \mp(1/4)\left((eR_g/2)^2 / |V(R_g)|\right) \times \mathcal{E}_0^2 f^2(t)\cos^2\theta$, provided $\omega_c \ll |V|/\hbar$. We are, however, concerned with the opposite extreme, $e\mathcal{E}_0 R_g \gg V_1(R_g)$, and thus take a different way to extract an effective rotational Hamiltonian, based on the general method of separating fast time scale from slow time scale in differential equations.[52,53] Following the approach developed for non-adiabatic localization in one-dimensional two-site systems,[41] we look for the solutions to Eq. (2) in the form: $\tilde{\mathbf{a}}(\mathbf{R},t) = \exp\left(i(e\mathcal{E}_0/(2\hbar\omega_c))s(t)(\hat{\mathbf{e}}\cdot\mathbf{R})\hat{\sigma}_z\right)\bar{\mathbf{a}}(\mathbf{R},t)$, where $s(t)$ is a fast-oscillating function: $s(t) = \omega_c \int_{-\infty}^{t} dt' f(t')\cos(\omega_c t') \approx \sin(\omega_c t) f(t)$, assuming $|d(\ln f(t))/dt| \ll \omega_c$, while $\bar{a}_1(\mathbf{R},t)$ and $\bar{a}_2(\mathbf{R},t)$ are supposed to be slower functions of $t$. Substitution of this expression in Eq. (2) leads, after some transformations, to the following system of equations in the spherical coordinates, whose polar axis is aligned with vector $\hat{\mathbf{e}}$:

$$i\hbar \frac{\partial \overline{\mathbf{a}}}{\partial t} = -\frac{\hbar^2}{M}\left(\frac{1}{R^2}\frac{\partial}{\partial R}\left(R^2 \frac{\partial}{\partial R}\right) + \frac{1}{R^2 \sin\theta}\frac{\partial}{\partial \theta}\left(\sin\theta \frac{\partial}{\partial \theta}\right) + \frac{1}{R^2 \sin^2\theta}\frac{\partial^2}{\partial \varphi^2}\right)\overline{\mathbf{a}} + V_c(R)\overline{\mathbf{a}} +$$

$$\frac{\hbar^2}{M}\left(\frac{1}{4}\left(\frac{e\mathcal{E}_0}{\hbar\omega_c}\right)^2 s^2 - is\frac{e\mathcal{E}_0}{\hbar\omega_c}\left(\cos(\theta)\frac{\partial}{\partial R} - \frac{\sin(\theta)}{R}\frac{\partial}{\partial \theta}\right)\right)\overline{\mathbf{a}} + V_1(R)\hat{\sigma}_x \exp\left(is\frac{e\mathcal{E}_0 R}{\hbar\omega_c}\cos(\theta)\hat{\sigma}_z\right)\overline{\mathbf{a}}. \quad (3)$$

In this equation, the second line concerns the time-dependent action of the laser field as manifested by the function $s(t)$. For the sake of simplicity, we assume at this point the strong rigid rotator approximation, in which the radial motion of nuclei is decoupled from the rotational degrees of freedom and is virtually unaffected by the single active electron, so that the ground and excited electronic states have the same radial nuclear wavefuncion. Consequently, we look for the solution of Eq. (3) in the form:

$$\overline{\mathbf{a}}(\mathbf{R},t) = \psi_R(R)\exp\left(-i\frac{E_R}{\hbar}t - \frac{ie^2\mathcal{E}_0^2}{4\hbar M\omega_c^2}\int_{-\infty}^{t}dt' s^2(t') + im\varphi\right)\overline{\overline{\mathbf{a}}}(\theta,t) \quad (4)$$

where $\psi_R(R)$ is the ground-state solution to the stationary radial Schrödinger equation, $E_R\psi_R = -(\hbar^2/M)(\partial^2\psi_R/\partial R^2 + (2/R)\partial\psi_R/\partial R) + V_c(R)\psi_R$. (Note that the second term in the argument of the exponential in Eq. (4) merely represents the very small ponderomotive energy shift of the heavy nuclei in the oscillating laser field.) Upon substitution of this $\overline{\mathbf{a}}(\mathbf{R},t)$ in Eqs. (3), radial averaging, and averaging with respect to $t$ over the laser cycle period $2\pi/\omega_c$ while neglecting the smaller terms of the order of $(e\mathcal{E}_0 R_g/(\hbar\omega_c))(1/(\omega_c\tau)) \ll 1$, we come to the following equation for the rotational motion, in which we use the conventional angular variable, $x = \cos(\theta)$:

$$i\hbar\frac{\partial \overline{\overline{\mathbf{a}}}}{\partial t} = -\frac{\hbar^2}{MR_g^2}\left((1-x^2)\frac{\partial^2}{\partial x^2} + 2x\frac{\partial}{\partial x} + \frac{m^2}{1-x^2}\right)\overline{\overline{\mathbf{a}}}(\theta,t) + V_1(R_g)J_0\left(\frac{e\mathcal{E}_0 R_g}{\hbar\omega_c}f(t)x\right)\hat{\sigma}_x\overline{\overline{\mathbf{a}}}(\theta,t) \quad (5)$$

where $R_g$ is the average value of $R$ in the ground state. (Strictly speaking, when applied to the terms in the right-hand side of Eq. (3) the angular averaging results in expressions, in which $R$ is replaced by an effective $R_g$, but these effective $R_g$s are slightly different for different terms. We disregard those small differences for the sake of notational simplicity.) The emergence of the Bessel function of zeroth order in the effective potential-energy term (the last term in the right-hand side of Eq. (5)) is typical in two-site tunneling suppression situations,[41] and it physically means that the electron charge becomes stuck in one of the sites. As seen in this term, now this nonadiabatic charge localization depends explicitly the molecular orientation through the $x$ variable. As the cycle-averaging of the terms with $\exp(is(e\mathcal{E}_0 R/\hbar\omega_c)\cos(\theta)\hat{\sigma}_z)$ in Eqs. (3) results in the same Bessel function of zeroth order in both of Eqs. (5), these latter equations are readily decoupled by the transformation $\mathbf{b}(x,t') = (\hat{\sigma}_z + \hat{\sigma}_x)\bar{\mathbf{a}}(x,t')$, so that the functions $\mathbf{b}(x,t')$ are determined by the dimensionless equations,

$$i\frac{\partial}{\partial t'}\mathbf{b}(x,t') = -\left((1-x^2)\frac{\partial^2}{\partial x^2} + 2x\frac{\partial}{\partial x} + \frac{m^2}{1-x^2}\right)\mathbf{b}(x,t') - vJ_0(\alpha f(t')x)\hat{\sigma}_z\mathbf{b}(x,t') \quad (6)$$

where the dimensionless time variable is scaled by the characteristic rotational time, $t' = t(MR_g^2/\hbar)$, $v = -V_1(R_g)(MR_g^2/\hbar^2) > 0$ determines the relative magnitude of the effective time-dependent potential, and $\alpha = (e\mathcal{E}_0 R_g)/(\hbar\omega_c)$ serves as the strong-field criterion parameter.

**Results and discussion**

In the case of a CW laser field ($f(t')=1$), the potential energy curves for the modified wavefunctions $b_1(x,t)$ and $b_2(x,t)$ in Eq. (6) depend on the field amplitude (through parameter $\alpha$), thus forming the surfaces presented in Figure 2. In this Figure, the energy is in the units of $MR_g^2 |V_1(R_g)|/\hbar^2$ and the electric field in the units of $\hbar\omega_c/eR$. The lower surface corresponds to the effective potential energy for $b_1(x,t)$; the upper surface corresponds to the effective potential energy for $b_2(x,t)$. It is instructive to compare these surfaces with those of Figure 1. As seen, the surfaces in Figure 2 have become corrugated, having additional extrema at intermediate values of the polar angle and the electric field strength. Moreover, even for moderately weak electric field, when the corrugation is not yet pronounced, the character of the extrema at $\theta = 0$ is reversed with respect to that in Figure 1. When the argument of the Bessel functions in Eqs. (6) is small, the effective potential energy in the first and the second equation behaves as $-v+(v/4)x^2\alpha^2 f^2(t')$ and $v-(v/4)x^2\alpha^2 f^2(t')$, respectively. This means that the effective polarizability of the molecule in the ground state is positive, and in the excited state is negative, in contrast to the case of a static electric field. The obvious physical reason for this weak-field reversal is that when $e\mathcal{E}_0 R_g \ll |V_1(R_g)|$, the time-averaging procedure leading to Eqs. (6) is conditioned on assumption $\hbar\omega_c > |V_1(R_g)|$, which implies the electronic response being out-of-phase with the driving laser field. As a result of this polarizability reversal, the equations do not revert to those for pendular states[54] even in the case of $\alpha \ll 1$. For instance, the action of the laser pulse on the ground-state molecule leads to anti-alignment rather than alignment of the molecular ensemble. Conversely, the action of the laser field on the excited-state molecules will result in alignment of the ensemble.

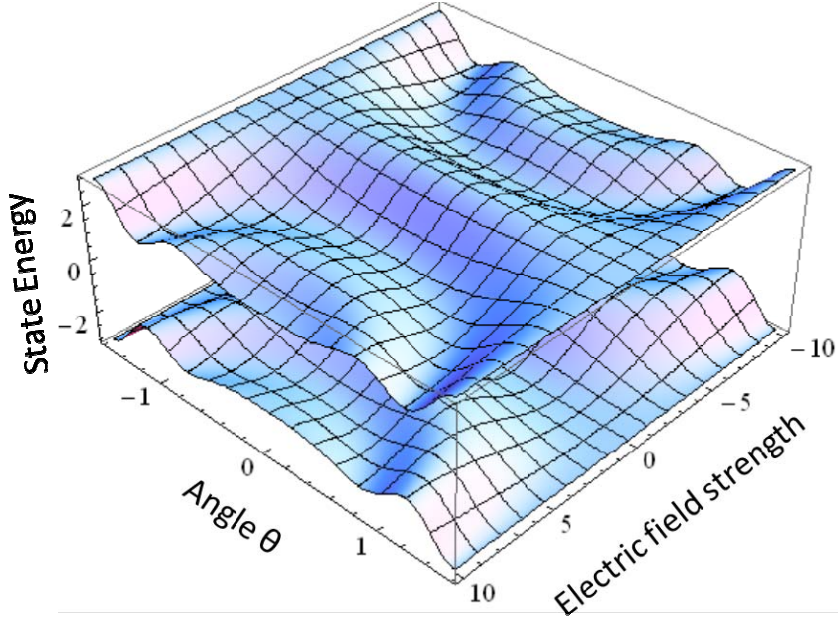

**Figure 2**. The effective potential energy curves of the cycle-averaged equations (Eqs. (6) in the text), as depending on the strength of the external electric field. The energy is in the units of $MR_g^2 |V_1(R_g)|/\hbar^2$, the electric field in the units of $\hbar\omega_c/eR$. The surfaces are shifted away from each other for clarity. Note the qualitative differences from the adiabatic case presented in Figure 1

In the case of an arbitrary pulse envelope $f(t')$, we look for the solutions of Eqs. (6) in the form of a series over normalized associated Legendre polynomials, $\tilde{P}_l^m(x,t) = P_l^m(x)\sqrt{(2l+1)/2}\sqrt{(l-m)!/(l+m)!}$, as $b_1(x,t') = \sum_l A_l^{(1)}(t') e^{-il(l+1)t'-ivt'} \tilde{P}_l^m(x)$, $b_2(x,t') = \sum_n A_n^{(2)}(t') e^{-in(n+1)t'+ivt'} \tilde{P}_n^m(x)$ (here, $l$ is the molecular angular momentum quantum number, the angular momentum being $L = \hbar l$). As equations (6) are decoupled, they can be considered separately. Then, for instance, the equation for $A_l^{(1)}$ is obtained as

$$\frac{\partial A_k^{(1)}}{\partial t'} = -iv\sum_l A_l^{(1)}(t') e^{it'(k-l)(k+l+1)} \int_{-1}^{1} dx \left(J_0(\alpha x f(t')) - 1\right) \tilde{P}_k^m(x) \tilde{P}_l^m(x) =$$

$$-iv\sum_l A_l^{(1)}(t') e^{it'(k-l)(k+l+1)} \sum_{n=1}^{\infty} \frac{(-1)^n}{(n!)^2} \left(\frac{\alpha}{2}\right)^{2n} (f(t'))^{2n} I_n^m(k,l),$$

(7)

where in the second line we use the series representation for the Bessel function and introduce the coefficients, $I_n^m(k,l) = I_n^m(l,k) = \int_{-1}^{1} dx\, x^{2n} \tilde{P}_k^m(x) \tilde{P}_l^m(x)$. These latter integrals can be calculated recursively, using the relation,

$$I_n^m(k,l) = \frac{1}{2k+3}\left(1 + 2\frac{k^2 - m^2}{2k-1}\right) I_{n-1}^m(k,l) +$$

$$\frac{1}{2k+3}\sqrt{\frac{\left((k+1)^2 - m^2\right)\left((k+2)^2 - m^2\right)}{(2k+1)(2k+5)}} I_{n-1}^m(k+2,l) + \quad (8)$$

$$\frac{1}{2k-1}\sqrt{\frac{\left(k^2 - m^2\right)\left((k-1)^2 - m^2\right)}{(2k+1)(2k-1)}} I_{n-1}^m(k-2,l),$$

obtained from the recurrence formula for the associated Legendre polynomials,[55] with $I_0^m(k,l) = \delta_{k,l}$.

As a proof-of-concept scenario, we consider how the rotational wavepacket is formed if the molecule is initially in the ground rotational state, $A_{k\,in}^{(1)} = \delta_{k0}$. In the case of a short laser pulse ($v\tau \ll 1$), the solution to Eq. (7) can be obtained iteratively. In particular, for a Gaussian laser pulse, $f(t') = \left(1/\sqrt{2\pi\tau^2}\right)\exp\left(-t'^2/(2\tau^2)\right)$, the composition of the resulting rotational wavepacket by the end of the pulse is found in the first approximation as,

$$A_{k\,fin}^{(1)} = \delta_{k0} - iv\tau\sqrt{\pi} \sum_{n=1}^{\infty} \frac{(-1)^n}{\sqrt{n}(n!)^2 (2\pi)^n} \left(\frac{\alpha}{2\tau}\right)^{2n} e^{-\frac{\tau^2}{4n}k^2(k+1)^2} I_n^0(k,0), \quad (9)$$

In this expression, the series converges very fast. As seen, the composition of the emerging wavepacket is mainly determined by the coefficients $I_n^m(k,l)$. Notably, these coefficients have nonzero albeit decreasing values for the values of $|k-l|$ ranging from 2 to $2n$, in a marked contrast with the traditional interaction, where only $|k-l|=2$ is allowed and thus the resulting wavepacket consists only of two states: $l=0$ and $l=2$. In contrast, the expression of Eq. (9) produces a rich wavepacket consisting of many states. For instance, for the parameters: $v=10^3$, $\tau=10^{-2}$, and $\alpha=10^{-2}$, corresponding to the laser pulse of 800 nm carrier wavelength, ~65 fs duration, and ~$10^9$ W/cm$^2$ intensity the resulting amplitudes are $A_{2\,fin}^{(1)}=-0.4696i$, $A_{4\,fin}^{(1)}=0.2615i$, $A_{6\,fin}^{(1)}=-0.1126i$, $A_{8\,fin}^{(1)}=0.0284i$, $A_{10\,fin}^{(1)}=-0.0045i$. As seen, the total contribution of states with $l=4$, $l=6$, and $l=8$ is almost as big as that of state with $l=2$, whereas only the latter state would emerge from the interaction in the case of polarizability-based rotational excitation, as was noted above. Note also that the sign of the $l=2$ contribution is flipped with respect to that in the polarizability-based case. This wavepacket enrichment, as well as the alignment-to-anti-alignment flip offers a direct means for experimental verification of the predicted new alignment regime.

The difference between the rotational wavepackets resulting from the traditional and the TNCR impulse interactions becomes even more pronounced when the electric field is strong enough to require non-perturbative treatment. In this situation, Eq. (7) was solved numerically, using variable-step fourth and fifth order Runge-Kutta method as implemented in the ode45 integrator in MATLAB. The system was initially in the ground rotational state, and the values of the dimensionless parameters were $v=10^3$, $\tau=10^{-2}$, and $\alpha=2.4$, corresponding to the pulse of 800 nm carrier wavelength, 65 fs duration, and $10^{13}$ W/cm$^2$ intensity. Using the same procedure, a numerical solution was obtained for the time-dependent rotational Schrödinger equation with the traditional interaction Hamiltonian, $\hat{H}_{int}=-(1/(4v))\left(Me^2 R_g^4 \mathcal{E}_0^2/\hbar^2\right) f^2(t)\cos^2\theta$, formally extended to the same values of parameters. The comparison of the after-pulse evolution of the resulting rotational wavepackets is presented in Figure 3. The presented curves trace the time dependence of the degree of alignment quantified as usual as $\langle \cos^2\theta \rangle$.[1] The gray dashed curve corresponds to the perturbational wavepacket produced by the traditional interaction. As was noted, in this case the wavepacket consists of just two states, $l=0$ and $l=2$, and the degree of alignment oscillates sinusoidally, with a small amplitude, around the isotropic value of 1/3. The cyan dash-dotted curve corresponds to the non-perturbative wavepacket resulting from the formally traditional interaction. As expected,[27] the shape of this curve substantially deviates from sinusoidal, and the average value is shifted upward of the baseline value of 1/3, indicating population transfer to higher rotational states. Finally, the red solid curve corresponds to TNCR interaction. This red curve is out-of-phase with the cyan curve (in the same manner as it was in the perturbational situation discussed in the previous paragraph), and the shape of this red curve is much richer than that of the cyan curve, indicating greater contribution of the states with higher rotational quantum numbers.

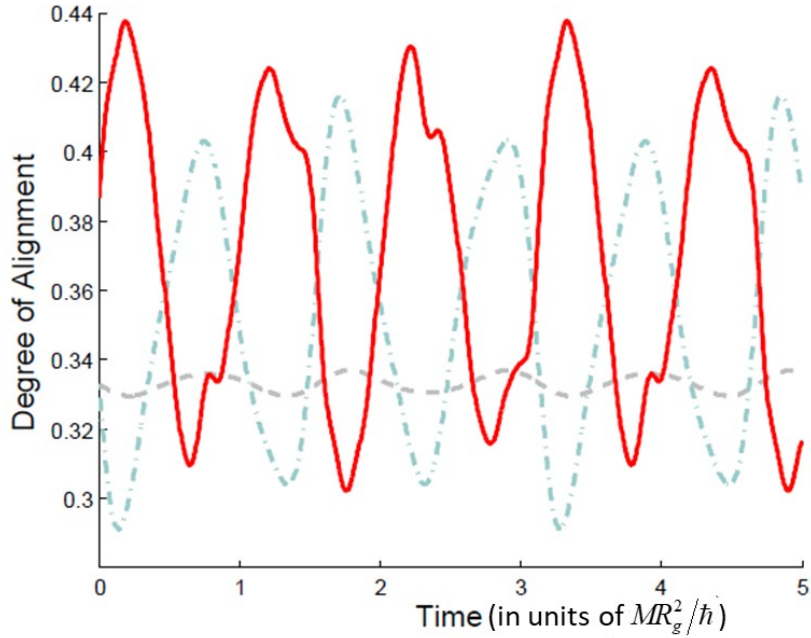

**Figure 3**. Evolution of the degree of molecular alignment (i.e., $\langle \cos^2\theta \rangle$) after the laser pulse. Grey dashed line: traditional interaction, perturbational regime; cyan dash-dotted line: traditional interaction, strong-field regime; red, solid line: TNCR interaction, strong-field regime (the dimensionless parameters in Eq. (6) being $\nu=10^3$ and $\alpha=2.4$, and the dimensionless pulse duration being $\tau=0.01$).

Finally, a few words are in order regarding experimental manifestations of the described effects. Although the considered conceptual model captures the essence of the phenomenon, one may expect the TNCR interaction mode to be manifested in real molecular systems in various degrees and with various complicating modifications. As seen from the foregoing discussion, the smaller the relative value of $V_1(R_g)$ in the model (that is, the smaller the interlevel energy distance in a real molecule), and the larger the value of the parameter $\alpha = (e\mathcal{E}_0 R_g)/(\hbar\omega_e)$ (that is, the larger the length of the molecule), the more likely the TNCR mode of interaction to be initiated. Using this as a guiding principle in choosing systems more suitable for experimental observation of TNCR effects, symmetric organic molecules of moderately small size may be likely candidates, such as polyenes (butadiene, hexatriene, and octatetraene) and polyacenes (naphthalene, anthracene, and tetracene), the upper size limitation dictated by the necessity to have the molecules in the gas phase. The length of these molecules ranges from ~ 5 Å to ~ 10 Å, and this makes for the value of $e\mathcal{E}_0 R_g$ ~ 5-9 eV at the laser intensity of $10^{13}$ W/cm$^2$, which is already greater than the energy gap separating the ground state from the excited states (typically, ~ 3-5 eV). In fact, however, mush lower laser intensities may suffice, based on the structure of excited state manifolds in these molecules. Although the ground state is separated from the excited state manifold by a sizable energy gap ~ 3-5 eV, the typical energy separation between the excited states is much smaller, of the order of 0.1 eV.[34,35,56,57] Thus, the excited molecules are going to favor TNCR mode of interaction with a typical near-IR laser pulse of 800 nm carrier wavelength ($\hbar\omega_c \approx 1.55$ eV) and moderate intensity ($\sim 5 \cdot 10^{11}$ W/cm$^2$). Moreover, when these molecules interact with a strong near-IR laser pulse, the nonresonant excitation proceeds through the so-called doorway state, the excited state for which the parameter $\Gamma = \left|\mu_{ge}\mathcal{E}_0 \hbar\omega_c\right|/(E_e - E_g)^2$ has maximum value (where $\mu_{ge}$ is the transition dipole matrix element from the ground state to the candidate excited state; $E_g$ and $E_e$ are the energies of these states). The calculations revealed that the doorway state is typically the lowest charge-transfer state.[34,35] When a molecule finds itself in this latter state, its interaction with the laser pulse and the resulting effective TNCR rotational kick can be well described by the model considered here. In this scenario, however, two factors are likely to complicate the expected results. First, the described nonresonant excitation by a linearly polarized laser pulse has naturally an angular dependence on its own. Second, the excited molecule is likely to continue gaining energy from during the laser pulse, resulting in ionization. Thus, for a proof-of-concept experiment, one might consider excitation of the molecule with UV pulse and interaction of the excited molecule with a moderate-intensity, non-ionizing near-IR pulse.

On the other hand, ionization of the molecule may pave the way for engaging THCR mechanism in the produced molecular ion. Indeed, the electron dynamics of large molecular ions in intense laser fields is different from that of neutral molecules, because in a ion there is a number of low-energy electronic transitions, corresponding to an electron hole migrating through the orbitals below the highest occupied molecular orbital (HOMO). Such nominally $\pi$-$\pi$, $\sigma$-$\pi$, and $\pi$-$\sigma$ transitions typically belong to the visible of near-IR range of the spectrum.[34,35] As a result, one can expect the TNCR kick mechanism to be engaged in a ground-state molecular ion. Then, the effective interaction of the intense laser pulse with the rotational degrees of freedom will comprise two TNCR stages, the first operating in the excited molecule and the second in the molecular ion.

**Conclusions**

A new mechanism for strong-field molecular alignment induced by impulsive interaction with an intense, linearly-polarized laser pulse is based on transient nonadiabatic charge redistribution in the electronic system of a molecule or molecular ion. This mode of electronic coupling with the oscillating laser field results in an effective interaction of the field with the molecular rotational degrees of freedom that is different from the traditional interaction Hamiltonian based on anisotropic polarizability. In turn, this alters the mechanism of rotational wavepacket formation and the patterns of subsequent alignment revivals in the molecular ensemble. This difference is clearly demonstrated in a simple case when the molecule is initially in the ground rotational state and interacts with a single short laser pulse. In this case, the rotational wavepackets that emerge from TNCR laser-molecule interaction are shown to contain much higher proportion of states with higher rotational quantum numbers, as compared to the wavepackets that would be produced via the traditional interaction mode based on anisotropic polarizability.

The proposed effects are modeled on a single-active-electron diatomic molecule in a tight-binding approximation. The effective rotational Hamiltonian is obtained by laser-cycle averaging and diagonalizing the time-dependent Schrödinger equation for the electronic and nuclear degrees of freedom. The electronically-nonadiabatic mode of effective interaction of strong-field laser pulses with the rotational degrees of freedom opens new ways for molecular alignment control based on different dependence of the nonadiabatic alignment kicks on the parameters of the molecular electronic system and on the laser pulse characteristics. The TCNR-type alignment kick mechanisms may be expected in cases when the strong laser field also causes considerable nonresonant excitation or ionization of a molecule.

## Acknowledgments

This work was supported by the National Science Foundation under Grants No. PHY-1806594 and CHE-0957694.